# High-responsivity MoS$_2$ hot-electron telecom-band photodetector integrated with microring resonator

Qiao Zhang, Yingke Ji, Siqi Hu, Zhiwen Li, Chen Li, Linpeng Gu, Ruijuan Tian, Jiachen Zhang, Liang Fang, Bijun Zhao[a)], Jianlin Zhao, Xuetao Gan[b)]

**AFFILIATIONS**

Key Laboratory of Light Field Manipulation and Information Acquisition, Ministry of Industry and Information Technology, and Shaanxi Key Laboratory of Optical Information Technology, School of Physical Science and Technology, Northwestern Polytechnical University, Xi'an 710129, China

[a)] Electronic mail: Bijun.zhao@nwpu.edu.cn

[b)] Author to whom correspondence should be addressed: xuetaogan@nwpu.edu.cn

**ABSTRACT**

We report a high-responsive hot-electron photodetector based on the integration of an Au-MoS$_2$ junction with a silicon nitride microring resonator (MRR) for detecting telecom-band light. The coupling of the evanescent field of the silicon nitride MRR with the Au-MoS$_2$ Schottky junction region enhances the hot-electron injection efficiency. The device exhibits a high responsivity of 154.6 mA W$^{-1}$ at the wavelength of 1516 nm, and the moderately uniform responsivities are obtained over the wavelength range of 1500 nm-1630 nm. This MRR-enhanced MoS$_2$ hot-electron photodetector offers possibilities for integrated optoelectronic systems.

During the past decade, two-dimensional (2D) transition metal dichalcogenides (TMDCs) have been widely studied owing to their unique properties, such as high carrier mobility,[1] dangling-bond-free surface,[2] and ideal flexibility,[3] which paved the way to co-integrate with a large variety of material systems such as silicon photonic integrated circuits without requiring crystal lattice matching.[4,5] However, most of the reported TMDCs-based photodetectors cannot be operated at the telecom-band due to their relatively large optical bandgaps. This limits their incorporation in the photonic integrated circuits and their further applications in telecom and datacom fields.

To address this, various hybrid structures have been proposed.[6-9] Among them, the simplest configuration is the metal-TMDCs heterostructure, showing hot-electrons excited by metallic nanostructures and strongly coupled with ultra-thin TMDCs. It provides an effective approach for enhancing light-harvesting in the telecom-band and photo-to-electron conversion efficiency.[10-13] In particular, few-layer $MoS_2$ functions as a favorable hot-electron acceptor as it has various traps at interfaces ($MoS_2$/metal, $MoS_2$/substrate) that trigger internal photogain.[6,14] Using these mechanisms, Park et al. reported a multilayer $MoS_2$ photodetector with responsivity reaching 0.539 mA $W^{-1}$ at a wavelength of 1550 nm by introducing plasmonic Ag nanocrystals.[15] Recently, our group demonstrated a $MoS_2$-based hot-electron photodetector integrated with an optical waveguide showing 15.7 mA $W^{-1}$ at a wavelength of 1550 nm.[16] However, their performance could be further improved by utilizing novel device structure/design that enhance the responsivity through increasing light absorption.

Herein, we propose an Au-$MoS_2$ hot-electron photodetector integrated with a silicon nitride microring resonator (MRR) to enhance light absorption and improve the photoresponsivity.[17] Using the MRR structure, the enhanced light-matter interaction makes it possible to realize light absorption efficiency comparable to a longer $MoS_2$ photodetector on the straight waveguide. At Au-$MoS_2$ interface, the evanescent field of the MRR excited at telecom-band wavelength would overlap with the surface plasmon in the Au film and hot electrons are generated via nonradiative decay. Then, these hot electrons are injected into the $MoS_2$ when their energies are greater than the Schottky barrier between Au and $MoS_2$, which could turn into photocurrents under an external bias across the $MoS_2$ channel. In addition, another electrode of the $MoS_2$ channel is designed with Ag, which would form a Ohmic contact with the $MoS_2$ layer and facilitate the carrier transport in the photodetector. With the fabricated device, we achieved a high responsivity of 154.6 mA $W^{-1}$ operating at 1516 nm and a fast photoresponse with rise/fall times of 9.1/11.3 μs, outperforming other works of waveguide-integrated $MoS_2$-based photodetectors.[18,19] We also demonstrated sub-bandgap photoresponse and a photocurrent spectrum that varies with the bias polarity, showing features governed by hot-electron injection. Our works provide possibilities to realize high photoresponsivity and high speed on-chip MoS$_2$-based optoelectronic devices at the telecom-band wavelength.

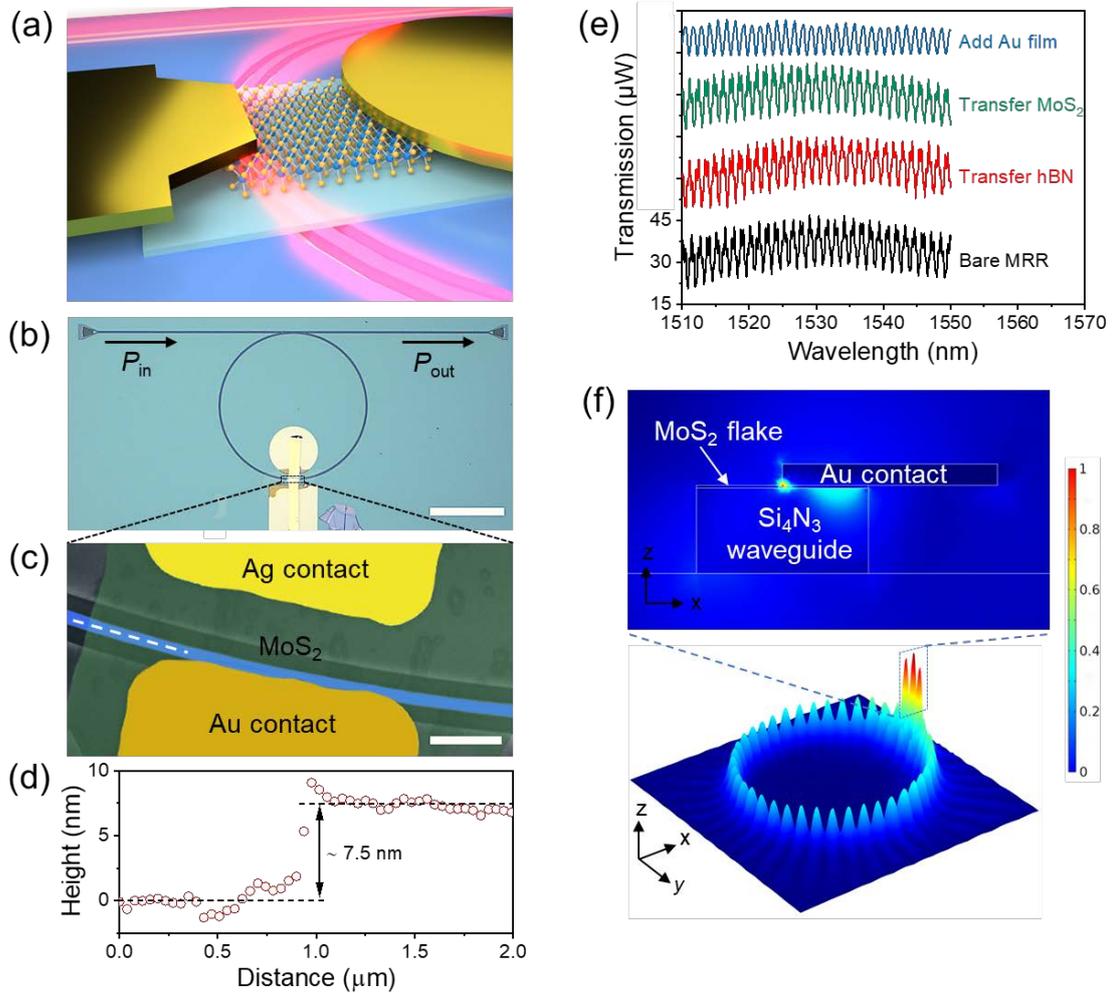

**FIG. 1.** (a) Schematic drawing of an MRR-enhanced MoS$_2$ hot-electron photodetector. (b) Optical micrograph of the device (top view) showing a few-layer MoS$_2$ flake integrated with a non-planarized silicon nitride MRR. Scale bar: 200 μm. (c) False-colour SEM image of a fabricated device, showing a silicon nitride MRR (blue), MoS$_2$ (green) contacted with Au contact (orange) and Ag contact (yellow) contacted. Scale bar: 10 μm. (d) Height profile of the MoS$_2$ flake examined by AFM, indicating a thickness of 7.5 nm. (e) Transmission spectra of the MRR measured after each step of fabrication. Spectra have been vertically translated for clarity. (f) Simulated electric field distribution in the silicon nitride MRR integrated with an Au-MoS$_2$ heterostructure,

where the MoS$_2$ flake across the strip waveguide of MRR and the Au electrode cover part of the strip waveguide.

The structure of the proposed device is illustrated in Fig. 1(a). The employed MRR were fabricated on a silicon wafer with a 300 nm thick top silicon nitride layer and a 3 μm thick buried oxide layer. The MRR has a radius, height and strip waveguide width of 150 μm, 300 nm and 1 μm, respectively.

Figure 1(b) displays the corresponding top-view optical microscope image of the fabricated device where the grating couplers at the two ends of the waveguide are used to couple with the optical signal from external light sources. Crystals of h-BN and MoS$_2$ were exfoliated onto the surface of polydimethylsiloxane (PDMS), using a Scotch-tape cleavage technique.[20-24] The final stack of 2D materials was aligned onto the waveguide chips using the micromechanical stage. The 2D layer stack consists of a few-layer MoS$_2$ flake (height ~7.5 nm, Fig. 1(d)) as the current-conduction semiconducting material, which is well aligned with the silicon nitride waveguide atop and directly contacted with a gold (Au) pad and a silver (Ag) pad on each side. The channel length of this MoS$_2$ photodetector is ~5 μm (with a waveguide width of ~1 μm), where the Au contact is positioned partly atop of the MRR to function as a light-absorbing metal material [Fig. 1(c)] and create photogenerated hot carriers. These carriers will be injected into the MoS$_2$ beneath and generate photocurrent across the channel. With the buried silicon oxide layer and the top silicon nitride layer as dielectric layers, a vertical electric field could be applied on the top MoS$_2$ layers when a gate voltage is applied between the Ag electrode and the bottom doped silicon substrate.

We optically characterize the fabricated MRR-integrated MoS$_2$ photodetector by coupling a wavelength tunable narrowband laser (TUNICS T100S-HP) into one of the grating couplers, and monitoring the output power at the other grating coupler. The coupling efficiency[25] could be expressed as $\eta = \sqrt{P_{fiber\_out} / P_{fiber\_in}}$, while $P_{fiber\_in}$ is the optical power in the fiber connecting source and silicon nitride chip, and $P_{fiber\_out}$ is the optical power in the fiber connecting silicon nitride chip and powermeter. The transmission spectra of the MRR were measured repeatedly by sweeping the laser wavelength from 1510 nm to 1550 nm for each step of fabrication [Fig. 1(e)]. Periodic interference fringes can be observed from the output spectra, which is determined by the phase difference between the bus waveguide and the MRR. Lorentzian fitting plot of the resonance curve shows resonance at ~1516 nm with a full width at half maximum (FWHM) of ~1 nm and a free spectrum range (FSR) of 1.3 nm of the MRR, which corresponds to a quality factor $Q$ (defined as the ratio of λ to FWHM) and finesse value (defined as the ratio of FSR to FWHM) of 1516 and 1.3, respectively. The improvement of coupling and the decrease of the $Q$ factor can be attributed to the shift of the MRR from the over coupled regime towards the critically coupled regime by inducing optical loss.[26]

We perform finite-element simulation on the MRR-integrated Au-$MoS_2$ structure to verify the enhancement of electric field and light absorption. As shown in Fig. 1(f), a whispering gallery mode is obtained in the MRR, which has an even stronger localized field over the region of the Au-$MoS_2$ junction. It could be attributed to the imaginary refractive index of the Au electrode. This strongly enhanced optical field around the Au electrode would promise large optical absorption and considerable generation of hot carriers.

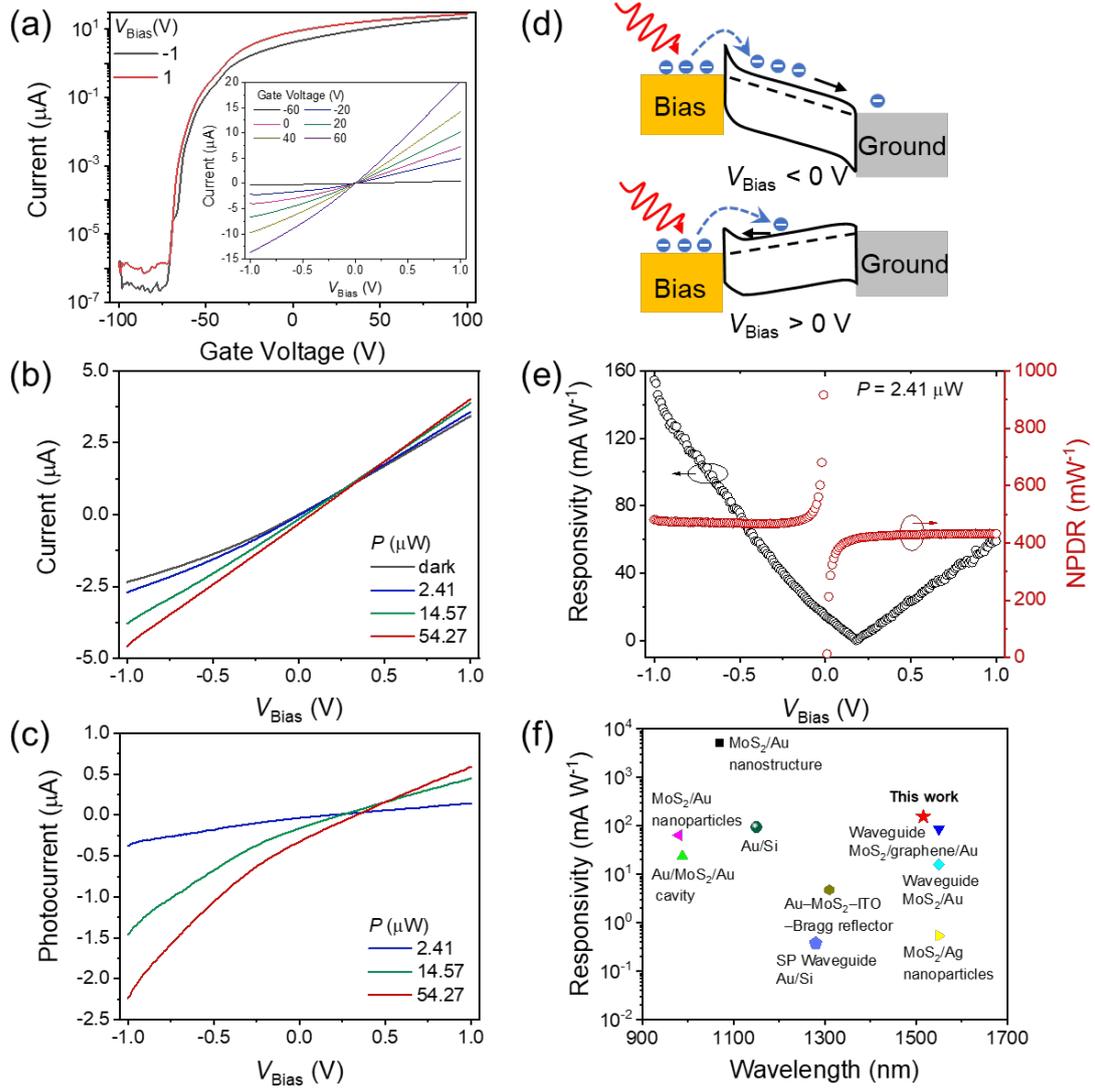

**FIG. 2.** (a) Electrical transfer characteristics of the MRR-integrated MoS$_2$ transistor, which are measured at $V_{Bias}$ = -1 V and 1V. Inset shows gate tunable current-voltage (*I-V*) characteristics of the photodetector. (b) Measured *I−V* characteristics of the photodetector at a linear scale without light (black line) and upon light incidence at various power intensities. (c) Photocurrent versus applied bias voltage of the photodetector at various incident optical powers. (d) Schematic energy band diagram explaining the photodetection mechanism under negative bias (upper diagram) and positive bias (lower diagram), respectively, showing the flow of hot-electrons with optical excitation. (e) Photoresponsivity and derived normalized photocurrent-to-dark-current ratio (NPDR) as a function of applied bias voltages. (f) Performance comparisons of our fabricated device with other hot-electron photodetectors.

We first performed field-effect transport measurements of the fabricated MRR-

integrated MoS$_2$ photodetector to characterize its electrical properties. The channel current, measured while sweeping the gate voltage at various fixed bias voltages, is shown in Fig. 2(a), indicating the behavior of an n-type transistor under the dark condition. The gate-tunable electrical characteristics were also investigated. The inset of Figure 2(a) presents the output current-voltage (*I-V*) characteristics under the dark condition, in which the current gradually increased at voltages between -1 and 1 V in accordance with an increase in the gate voltage from -60 to 60 V. The slight nonlinear behavior in the output *I-V* characteristics can be attributed to the Schottky barrier formation between Au and MoS$_2$ and the asymmetric structure formation between Au and Ag pads.

Figure 2(b) shows the representative *I-V* characteristics of the Au/MoS$_2$/Ag photodetector with asymmetric metal contact under illumination. The laser with wavelength at 1516 nm is directed into the grating coupler, guided along the waveguide, and absorbed at the Schottky junction, where a photocurrent is generated. The *I-V* curves show similar characteristics with varied illumination powers. The magnitude of the photocurrent increases with optical power that more photons are absorbed, more photogenerated carriers are generated and collected. The generated photocurrent, $I_{\text{ph}}$, can be estimated with the following formula

$$I_{\text{ph}} = I_{\text{illumination}} - I_{\text{dark}}$$

Derived from the *I-V* curves, Figure 2(c) shows the photocurrents as a function of the applied bias voltage at different incident powers. The photocurrent strongly increases at the reverse bias voltage region as an expected behavior for hot-electron photodetector.[27]

To shed light on the operation mechanism of the hot-electron-based photodetector, simplified energy band diagrams are employed to describe the photocurrent generation process taken place under illumination as depicted in Fig. 2(d). The Ag electrode is set as ground, while the bias voltage is applied to the other electrode. Because the work function of Au metal (5.1 eV) is higher than that of the MoS$_2$ (4.0 eV), the electrons in MoS$_2$ would be injected into the Au electrode in order to maintain the equilibrium of Fermi energy when Au and MoS$_2$ contacted with each other. In this way, an ideal built-in electric field of 0.5 eV could be formed in principle which can suppress the dark current at zero bias voltage.[28] The Ag electrode away from the waveguide has a relative low work function of 4.26 eV, which means this Ag electrode formed an almost perfect Ohmic contact with the intrinsic n-type MoS$_2$ channel. Notably, the discussion above ignores the influences of any interface state, which relates to Fermi level pinning and effective work function reduction of both Au and Ag.[29] Meanwhile, due to the low barrier of the Au−MoS$_2$ junction, the hot electrons generated by the Au electrode would transfer into the conduction band of MoS$_2$ before their thermalization. At $V_{Bias} < 0$ V within the breakdown voltage (upper diagram of Fig. 2(d)) the strength of the net electric field of the device is enhanced, which comparatively accelerate the hot electrons injected from the Au electrode across the MoS$_2$ channel to the Ag electrode. In contrast, as shown in the lower diagram of the Fig. 2(d), at $V_{Bias} > 0$ V the electric field would suppress the injection process of hot electrons from the bias electrode to the MoS$_2$ channel. Therefore, the Au-MoS$_2$ Schottky junction is confirmed to play a crucial role in the photodetection process studied in this work, in which this barrier can be modulated precisely by the bias voltage for different operation modes of the detector.

As shown in Fig. 2(e), the photoresponsivity increases with the applied bias voltage, as more carriers can be extracted before recombination. Note that the corresponding external photoresponsivity ($R$) of the photodetector is defined as the ratio of the photocurrent to the incident illumination power (that is, $I_{ph}/P_{in}$), where $P_{in} = P_{fiber\_in} * \eta * (1-e^{-\alpha L})$ with $\alpha$ is the absorption coefficient and $L$ is the width of the transferred Au-MoS$_2$ junction.[30] The responsivity varies linearly as a function of bias voltage and shows that the device is not yet driven into saturable absorption at these power levels. The derived responsivity under -1 V bias is 154.6 mA W$^{-1}$ at a wavelength of 1516 nm, which is higher than the reported waveguide-integrated hot-electron MoS$_2$ photodetector.[16]

To quantify the sensitivity of a photodetector integrated on a waveguide, the normalized photocurrent-to-dark-current ratio (NPDR) is an important metric. As a result of the very low dark current while taking into account the extracted photoresponsivity, the NPDR of our device is remarkably high. As presented in Fig. 3(d), our device achieves an NPDR of 300-500 mW$^{-1}$ in forward bias and up to 600 mW$^{-1}$ in reverse bias, which is several orders of magnitude higher than those of, for example, graphene photodetectors.[31] These values exceed the reported NPDR for waveguide-integrated BP photodetector[32] and are on par with MoTe$_2$ waveguide photodetector.[33] These values underline the high photodetection efficiency of our device, considering that only a fraction of the light is absorbed by evanescent coupling.

To justify the performance of the realized MRR-integrated MoS$_2$ photodetector at the telecom-band wavelength, the responsivity and detection wavelength range of the photodetector and other reported devices are compared and shown in Fig. 2(f).[15,16,18,34-39] These comparisons have proven that the MoS$_2$-based hot-electron photodetector constructed on the MRR possesses excellent electrical and optoelectronic properties. The excellent performance of our device is mainly arisen from (1) the hot-electron transfer process and (2) the photon lifetime enhancement in the MRR, which is proportional to the finesse of the cavity.

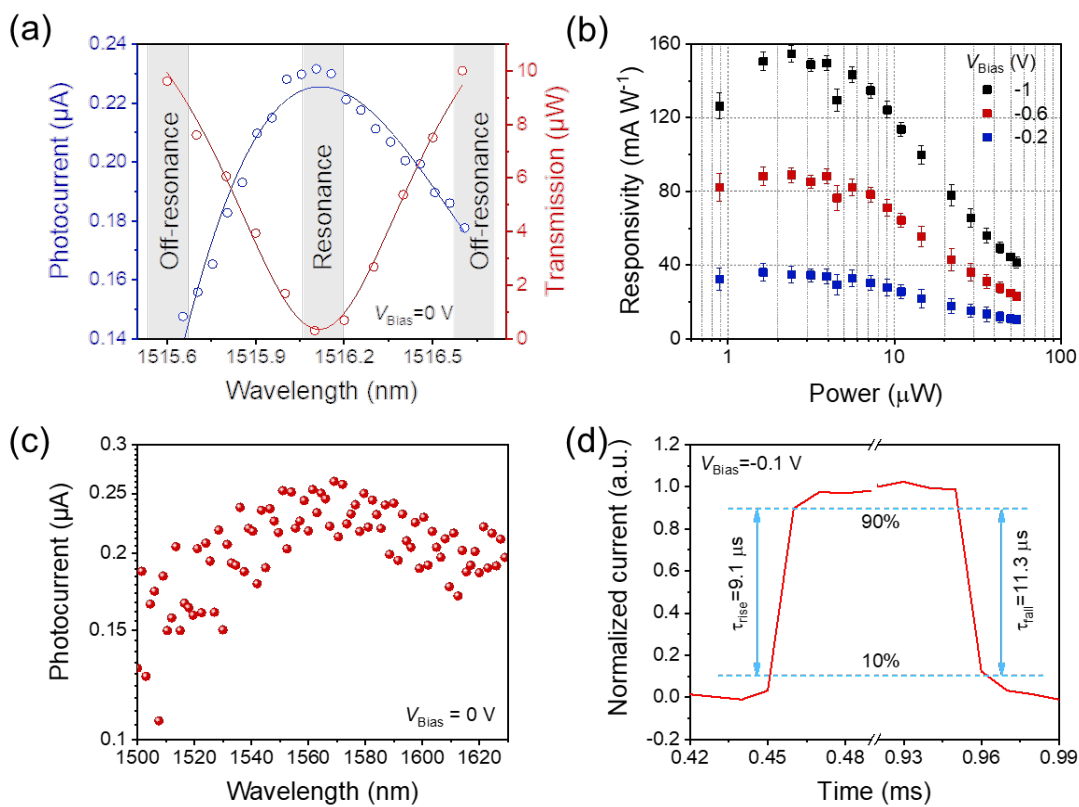

**FIG. 3.** (a) Spectral response of the MRR-integrated MoS$_2$ photodetector, showing

maximum responsivity at the resonance wavelength (1516.1 nm) for the bias of -1 V (open blue spheres) when the optical transmission (open red spheres) is at its minimum. (b) Responsivity of the MMR-integrated $MoS_2$ photodetector as a function of illuminated optical power for -1 V, -0.6 V and -0.2 V, respectively. (c) Spectral response of the MRR-integrated $MoS_2$ photodetector at a wavelength ranging from 1500 to 1630 nm. (d) Response time of the $MoS_2$ photodetector under the excitation of the modulated laser.

We then consider the impact of the MRR on the photodetector's performance. When operated at a fixed power of 54.27 μW at zero bias, the photocurrent exhibits more than 57% enhancement at the resonance wavelength compared with that at off-resonance wavelength, which overtakes the MRR's finesse due to the reinforcement of hot electrons [Fig. 3(a)]. It should be noted that the photon lifetime of the MRR needs to trade off with the 3dB response, although a higher fineness can enhance the responsivity. In addition, such an MRR-enhanced photodetector is wavelength sensitive, which plays an important role in spectrally resolved applications such as wavelength division multiplexing systems.[40,41]

The measured responsivity shows a downward trend with higher optical input power and reaches a flat response until $P_{input}$ = 6 μW. This is because under higher input power, a larger number of hot electrons would be generated at the Au/$MoS_2$ interface leading to an enlarged transfer rate of hot electrons. Therefore, negative electrons will the accumulated in the $MoS_2$, while positive charges accumulate in the Au electrode, which creates an internal electrostatic field and blocks the hot electrons from leaving the Au pad [Fig. 3(b)].[42] The spectral response of the MRR-integrated photodetector ranging from 1500 to 1630 nm is shown in Fig. 3(c), where the wavelength-dependent photoresponse is observed. The photoswitching characteristics were also measured by modulating a tunable laser at a frequency of 1 KHz. The normalized transient responses for the modulated light pulse are shown in Fig. 3(d). The rise time and fall time are defined as the time between 10% and 90% of the current at leading and falling edges, and the estimated values are 9.1 and 11.3 μs, respectively. The fast rise and fall times benefit from the fast carrier mobility of $MoS_2$ and the short channel length between the

Au and Ag pads. To summarize, this MRR-integrated photodetector is beneficial to improve the responsivity of the $MoS_2$-based photodetector and its wavelength-sensitive properties make it potential for applications in communication, spectroscopy and sensing in wavelength-division multiplexing systems.

In conclusion, we have demonstrated a $MoS_2$-based hot-electron photodetector integrated on a silicon nitride MRR, which has a high photoresponsivity at the telecom-band wavelength. By integrating an Au-$MoS_2$ Schottky junction partly on the strip waveguide of the MRR, its coupling with the evanescent field of the MRR could generate considerable hot electrons that could be injected into the $MoS_2$ channel and yield photocurrent. A responsivity of 154.6 mA·$W^{-1}$ at 1516 nm (applied voltage of -1 V) and rise/decay times of 9.1/11.3 μs are obtained from the fabricated device. The enhancement of the photodetection by the MRR mode shows a 57% improvement in photocurrent at the resonance wavelength comparing with that at the off-resonance wavelength. Our results have demonstrated the possibilities to integrate $MoS_2$ photodetectors with silicon nitride photonic platform, which paves the way for more extensive applications in spectroscopy, sensing, and communication fields.

## SUPPLEMENTARY MATERIAL

See supplementary material for the tunable optoelectronic characteristics of the MRR-integrated $MoS_2$ photodetector by simple electrostatic gatings and other MRR-integrated $MoS_2$ photodetector.

## ACKNOWLEDGMENTS

We gratefully acknowledge financial support from the Key Research and Development Program (Grant Nos. 2018YFA0307200), the National Natural Science Foundations of China (Grant Nos. 61905196, 62105267), the Key Research and Development Program in Shaanxi Province of China (Grant Nos. 2020JZ-10). The authors thank the Analytical & Testing Center of NPU for the assistance of device fabrication.

## AUTHOR DECLARATIONS

### Conflict of Interest

The authors declare no conflicts of interest.

## DATA AVAILABILITY

The data that support the findings of this study are available from the corresponding author upon reasonable request.